\documentclass[amsmath,amssymb,aps,pre,reprint]{revtex4-1}

\usepackage{graphicx}
\usepackage{bm}
\usepackage{amsthm}

\newtheorem{corollary}{Corollary}

\newcommand{\n}{\noindent}
\newcommand{\R}{\mathbb{R}}
\newcommand{\E}{\mathbb{E}}

\newcommand{\Z}{\mathbb{Z}}

\begin{document}
\preprint{APS/123-QED}

\title{Noise Sharing and Mexican Hat Coupling in a  Stochastic Neural Field}
\author{Peter H. Baxendale$^1$}
\address{$^1$Department of Mathematics, University of Southern California, Los Angeles, CA, USA}
\author{Priscilla E. Greenwood$^2$}
\address{$^2$Department of Mathematics, University of British Columbia, Vancouver, BC, Canada}
\author{Lawrence M. Ward$^3$} 
\address{$^3$Department of Psychology and Brain Research Centre, 2136 West Mall, University of British Columbia, Vancouver, BC, V6T 1Z4 Canada}
\thanks{Corresponding author. Email: lward@psych.ubc.ca,  Tel.: +1 604 822 6309, Fax: +1 604 822 6923.}
\date{\today}
 
\begin{abstract}
\noindent 
 A diffusion-type coupling operator biologically significant in neuroscience is a difference of Gaussian functions (Mexican Hat operator) used as a spatial-convolution kernel. We are interested in pattern formation by \emph{stochastic} neural field equations, a class of space-time stochastic differential-integral equations using the Mexican Hat kernel. We explore, quantitatively, how the parameters that control the shape of the coupling kernel, coupling strength, and aspects of spatially-smoothed space-time noise, influence the pattern in the resulting evolving random field. We confirm that a spatial pattern that is damped in time in a deterministic system may be sustained and amplified by stochasticity. We find that spatially-smoothed noise alone causes pattern formation even without direct spatial coupling. Our analysis of the interaction between coupling and noise sharing allows us to determine parameter combinations that are optimal for the formation of spatial pattern.

\vspace*{2ex}\noindent\textit{\bf Keywords}: neural field equation, Mexican Hat coupling, stochastic process, spatial pattern, spatially-smoothed noise.
\\[3pt]
\\[3pt]
\end{abstract}

\maketitle


\section{\label{intro}Introduction}
In this paper we explore the formation of pattern by a stochastic neural field equation with simple damping as its reaction term  and with both i.i.d. and shared noise. The existence and possible sources of shared noise correlations in stochastic neural models have been studied recently by Doiron et al. \cite{Doiron16}, and by Meyer, Ladenbauer, and Obermayer \cite{Meyer17}. We, on the other hand focus on the effect of shared or correlated noise on pattern formation. 

We have been shown by Hutt and colleagues \cite{Hutt2003} that spatial pattern, embedded in deterministic space-time dynamics but immediately damped, may be excited by noise. Butler and Goldenfeld \cite{Butler09, Butler11}, and McKane, Biancalani and Rogers \cite{McKane14} showed the existence of excitable spatial modes that, when noise was added, were revealed in power spectral densities. The knowledge of this noise-facilitated source of pattern, also observed in biological systems, motivated us to explore how certain sample path properties of evolving stochastic neural fields depend on the parameters in a basic example. We look at parameters that control strength of coupling or local interaction in the field, and on the extent of local sharing, or smoothing, of noise. 

The sample path properties we look at are: how pattern grows with coupling strength, how pattern is revealed and sustained by noise when field interaction has no excitable modes, how coupling interacts with noise smoothing, giving rise to distinct patterns for optimum smoothing, and then yielding distorted pattern with "too much smoothing."  A striking result is that spatial smoothing of noise, alone, without direct neural field interaction, produces pattern.

Simulation of the time evolution of one-dimensional fields yields insight about typical sample path behaviour of the evolving random field and illustrates the analytical conclusions we present. We use two measures of spatial pattern. One is the spatial FFT amplitude as a stochastic process in time (note, for comparison to other studies, that power spectral density is the square of the FFT amplitude; here we compute spatial FFT amplitudes directly and from simulations). The second is a function of space, which we call $F$, that allows direct observation of dominant frequency, again a stochastic process in time. We believe that display of sample paths of both of these processes is innovative here, and that they will prove to be valuable in future studies of the model we study and of other stochastic neural fields. 

What has come to be called a \emph{neural field equation} is an integro-differential equation of the form
\begin{equation}\label{nfe}
dY(t, x) = \bigg[-Y (t, x) +\int_{\mathbb{R}}cw(x-y)S(Y (t, y))dy\bigg]dt,
\end{equation}
where $Y$ is an $\mathbb{R}^1$-valued state variable for a neural system, $w$ is a coupling operator, for example the Mexican Hat convolution kernel, $c$, is a constant called the coupling strength, and $S$ is a scaling functional, typically a sigmoid, which keeps $Y$ bounded in the diffusion term. In what follows we take $S$ to be the identity within a certain bounding range. We study a stochastic version of (\ref{nfe}),  wrapped with length $L$, 
\begin{equation}\label{nfes}
\begin{split}
&dY(t, x) = \bigg[-Y (t, x) +\int_{0}^{L} c w(x-y)Y (t, y)dy\bigg]dt\\
&+\sigma dG(t, x),
\end{split}
\end{equation}
where $\sigma$ is a constant diffusion coefficient, $c\geq 0$ is a constant, and $G(t,x)$ is a Gaussian process that may depend on both time, $t$, and space, $x$. We study the interaction of coupling and noise smoothing in \eqref{nfes} theoretically in continuous space and time, and then discretize both for simulations. 

Recently, the study of such equations, with continuous state variable, has been put on a rigorous footing by Faugeras and Inglis \cite{Faug15}. Conveniently for the study of neural field equations that generate spatial pattern, they singled out the difference-of-Gaussians coupling operator (often called the Mexican Hat operator) as one that satisfies established conditions for the existence and uniqueness of a solution. As will be seen, this coupling operator generates pattern when used in equations like (\ref{nfes}). It is sometimes preferred to the Laplacian as a coupling operator \cite{Scars11}. 

In what follows we compute conditions for the interaction of the dominant modes produced by coupling and noise smoothing in the stochastic model (\ref{nfes}). When no excitable modes are present the spatial pattern generated by the dominant modes is damped, but added noise reveals and sustains the damped spatial pattern. Noise smoothing can either render the revealed pattern more clearly, or distort it, depending on the extent of the smoothing. 

New here is an analytic and graphical study of solutions of \eqref{nfes} with smoothed noise. We are interested in spatial pattern revealed by noise in stochastic sample paths from the evolving random field of the neural field equation as parameters are varied. We explore the relation between the coupling strength constant, $c$, and a parameter of noise smoothing, $\eta$, the standard deviation (SD) of a Gaussian smoothing kernel. We consider separately the ranges of $c$ where the eigenvalues are all negative and where the maximum eigenvalue is positive. In our computations a specific value of $c$ separates these ranges.  Also new is the result that smoothed noise, by itself with $c=0$ in \eqref{nfes}, can produce spatial patterns similar to those arising from excitable spatial modes created by the Mexican Hat coupling.
 
\section{\label{stnoise}Input noise}
In modeling a spatially discrete stochastic neural field, a default choice (e.g., \cite{Touboul14}) has been to introduce an independent Brownian component for each $x$. If the locations are tightly packed, however, as in neural tissue, the same input noise may be shared in a neighborhood of locations. This is because the nearer neurons are to each other in the brain the more similar is the set of synaptic inputs they receive. Neurons communicate more extensively with each other the closer they are to each other because the number and strength of synaptic connections between neighboring neurons tends to decline with the distance between them \cite{Song05, Haider09}. Given that shared noise would be mostly synaptic noise \cite{Haider09, Faisal08}, the amount of shared noise between two neurons should decline with the amount of effective connectivity between them. We are interested in the effect on spatial patterns of the size of the neighborhood in which some of the same input noise is felt. In our exploration we have included cases where noise is independent at each location, $x$, and cases where noise is averaged over neighborhoods of various sizes. We refer to this as spatial smoothing of noise. It has been shown that such noise allows for (H\"older) continuous solutions to a broad class of stochastic integro-differential equations, including equations such as (\ref{nfes}) \cite{Dalang98, Faug15, Ferrante06, Sanz02}. In simulations, Section \ref{sims}, for spatial smoothing we used a Gaussian kernel. The Gaussian smoothing kernel was convolved with i.i.d. Gaussian noises at each iteration of the evolving spatial field. 

\section{\label{nfeq}Interaction of Coupling and Noise Smoothing in The Neural Field Equation }
We capture the interaction of coupling and noise smoothing in terms of the Fourier component processes associated with \eqref{nfes}. We begin with \eqref{nfes}, taking $w$ and $r$ to be general functions.  Later we specialize to Mexican Hat and Gaussian functions.  The noise $G(x,t)$ is Gaussian with mean 0 and covariance
  \begin{equation}\label{Gnoise}
     \E\left[G(t,x)G(s,y)\right] = \min(s,t) r(x-y).
\end{equation}
Here $r$ is a general positive semi-definite function. We denote the Fourier transforms of the coupling kernel $w$ and the noise smoothing function $r$ by
    $$
    W(k) = \int_{-\infty}^\infty e^{-ikx} w(x)\,dx, \quad k \in\R,
        $$
and
     $$
    R(k) = \int_{-\infty}^\infty e^{-ikx} r(x)\,dx, \quad k \in\R.
    $$

The spatial variable $x$ is in an interval $[0,L]$ with periodic boundary conditions, equivalently on the circle $\R/L\mathbb{Z}$, so long as the functions $w$ and $r$ are made periodic.  That is, replace $w(x)$ by $\hat{w}(x) = \sum_{n \in\mathbb{Z}} w(x+nL)$ and $\hat{r}(x) = \sum_{n \in\Z} r(x+nL)$ and drop the `hat' notation. In fact the widths of the support of $r$ and $w$ will be less than $L$.

The periodic function $w(x)$ has Fourier coefficients
  \begin{equation}\label{WkF}
   W(2 \pi k/L) = \int_0^L e^{-2 \pi ikx/L}w(x)\,dx, \quad k \in \mathbb{Z}, 
        \end{equation}
and the periodic function $r(x)$ has Fourier coefficients
  \begin{equation}\label{RkF}
    R(2 \pi k/L) = \int_0^L e^{-2 \pi ikx/L}r(x)\,dx, \quad k \in \mathbb{Z}.
     \end{equation}
Moreover $r$ is positive semi-definite. 
We consider the model \eqref{nfes}, where the noise $G(x,t)$ is given by \eqref{Gnoise}. A Fourier series expansion of the solution $Y(t,x)$ of \eqref{nfes} allows us to write it in terms of a family of Ornstein-Uhlenbeck processes indexed by spatial frequencies (wave numbers, $k$). We can then compute the expected squared amplitudes of these O-U processes.
     
{\bf Proposition} (cf.~\cite{Hutt08}). There are standard 2-dimensional Brownian motions $\{C_k(t): t \ge 0\}$ for $k \ge 1$ and a standard scalar Brownian motion $\{C_0(t): t \ge 0\}$, all mutually independent, such that the solution $Y(t,x)$ of \eqref{nfes} is given by
     \begin{equation} \label{YFour}
     Y(t,x) = a_0(t) + \sum_{k \ge 1} 2{\rm Re}\left(a_k(t) e^{2 \pi i k x/L}\right)
     \end{equation}
where the scalar process $a_0(t)$ satisfies
     \begin{equation} \label{a0}
     da_0(t) = \left[-1+ cW(0)\right]a_0(t)\,dt + \frac{\sigma}{\sqrt{L}}\sqrt{R(0)}\,dC_0(t)
     \end{equation}
and for $k \ge 1$ the complex processes $a_k(t)$ satisfy
    \begin{equation} \label{ak}
    \begin{split}
     &da_k(t) = \left[-1+ cW(2\pi k/L)\right] a_k(t)\,dt\\
     &+ \frac{\sigma}{\sqrt{2L}}\sqrt{R(2 \pi k/L)}\,dC_k(t).
     \end{split}
     \end{equation}

\begin{proof} Substituting the Fourier series expansion
   $$
    Y(t,x) = \sum_{k \in \mathbb{Z}} a_k(t)e^{2 \pi ikx/L}
    $$
  into \eqref{nfes} gives
  \begin{equation*}
  \begin{split}
  &\sum_{k \in \Z} da_k(t) e^{2 \pi ikx/L} =\\ 
  &-\sum_{k \in \Z} a_k(t)e^{2 \pi ikx/L}dt\\
  &+ c\int_0^L w(x-y) \sum_{k \in \Z} a_k(t)e^{2 \pi iky/L}dydt\\
  &+\sigma dG(t,x).
  \end{split}
  \end{equation*}
Since
  \begin{equation*}
  \begin{split}
&  \int_0^L w(x-y) e^{2 \pi iky/L}\,dy \\
&= e^{2 \pi ikx/L}  \int_0^L w(x-y) e^{2 \pi ik(y-x)/L}\,dy\\
&  = e^{2 \pi ikx/L}W(2 \pi k/L)
\end{split}
\end{equation*}
we get
 \begin{equation*}
 \begin{split}
  &\sum_{k \in \Z} da_k(t) e^{2 \pi ikx/L} \\
  &=\sum_{k \in \Z}\bigl[- 1+ c W(2 \pi k/L)\bigr] a_k(t)e^{2 \pi ikx/L}\,dt+\sigma dG(t,x).
  \end{split}
  \end{equation*}
Now multiply by $e^{-2 \pi i \ell x/L}$ and integrate with respect to $x$.  We get
  $$
   L  da_\ell(t)  = L \left[-1+  c W(2 \pi \ell/L)\right] a_{\ell}(t)\,dt+\sigma dB_\ell(t),
   $$
equivalently
    \begin{equation} \label{akk}
  da_k(t)  =  \left[-1+  c W(2 \pi k/L)\right] a_k(t)\,dt+\frac{\sigma}{L} dB_k(t)
  \end{equation}
where
 \begin{equation} \label{Bk}
    B_k(t) = \int_0^L e^{-2 \pi i k x/L} G(t,x)\,dx, \quad k \in \Z.
  \end{equation}
Assuming that the initial condition $Y(0,x)$ is real, we have $a_{-k}(0) = \overline{a_k(0)}$.  Since $B_{-k}(t) = \overline{B_k(t)}$ for all $t$, we have $a_{-k}(t) = \overline{a_k(t)}$ for all $t \ge 0$.  Therefore $a_k(t)e^{2 \pi i kx/L}+a_{-k}(t)e^{-2\pi i k x/L} = 2\mbox{Re}\left(a_k(t)e^{2 \pi i kx/L}\right)$, and it suffices to study $a_k(t)$ for $k \ge 0$.

 Using the calculations in Appendix A, we may write $B_0(t) = \sqrt{L R(0)}C_0(t)$ and $B_k(t) = \sqrt{(L/2)R(2 \pi k/L)}C_k(t)$ for $k \ge 1$, where the processes $C_k$ are as stated in the proposition.  For $k \ge 1$ the equation \eqref{akk} can then be rewritten as \eqref{a0} and \eqref{ak}.      \qedhere
\end{proof}

\begin{corollary}  Suppose
    $$
    Y(0,x) = A_0 + 2\sum_{k \ge 1} A_k \cos \bigl(2 \pi k x/L + \phi_k\bigr)
    $$
Let $a_0(t)$ be the solution of \eqref{a0} with initial condition $a_0(0) = A_0$, and for $k \ge 1$ let $a_k(t)$ be the solution of \eqref{ak} with initial condition $a_k(0) = A_ke^{i \phi_k}$.  Write $a_k(t) = A_k(t)e^{i \phi_k(t)}$.  Then
     $$
    Y(t,x) = a_0(t) + 2\sum_{k \ge 1} A_k(t) \cos \bigl(2 \pi k x/L + \phi_k(t)\bigr).
    $$
\end{corollary}

Thus the period $k$ part of $Y(t,\cdot)$ has amplitude $2A_k(t)$ and phase $\phi_k(t)$ determined by the Ornstein-Uhlenbeck process $a_k$.

Look more closely at the process $a_k(t)$.  We consider $k \ge 1$ (the $k = 0$ case is similar).   For ease of notation write \eqref{ak} as
     $$
     da_k(t) = \lambda_k a_k(t)\,dt + \sigma_k \,dC_k(t).
     $$
This complex-valued SDE has solution
    \begin{equation}\label{akt}
    \begin{split}
    &a_k(t) = e^{\lambda_k t}a_k(0) + \sigma_k \int_0^t e^{\lambda_k(t-s)} dC_k(s)\\
    & \equiv e^{\lambda_k t}a_k(0) + N_k(t),
   \end{split}
   \end{equation}
   \cite{Gard0x}.  Here 
   $$\lambda_k = -1+ cW(2 \pi k/L)$$
    and 
    $$\sigma_k = \frac{\sigma}{\sqrt{2L}}\sqrt{R(2 \pi k/L)}$$
     for $k \ge 1$ (and slight modifications to the formulas if $k = 0$). 
     
We can now outline our overall strategy. Because the integral \eqref{nfes} is over a bounded set, the eigenvalues of the operator in \eqref{nfes} are separated, and the mode $k$ of the dominant eigenvalue determines the spatial frequency of the spatial pattern. For each pair of parameters $(c,\eta)$, which will determine the coupling strength and the width of the noise smoothing, defined in section \ref{specnoise}, we are able to evaluate a functional of the distribution of the process $Y_{c,\eta}$ that is maximal at that value of $(c,\eta)$ for which $k_{c,\eta}$ is the dominant mode. This functional, the expected squared amplitude of the $k$th mode of $Y_{c,\eta}$, can be computed in terms of the drift and diffusion coefficients, $\lambda_k, \sigma_k$, of the process $a_k (t)$ defined by \eqref{a0}, \eqref{ak}. 

Since $C_k$ is standard 2-dimensional Brownian motion, then $N_k(t)$ is 2-dimensional Gaussian with mean zero and covariance matrix 
   $$
  \sigma_k^2 \left( \int_0^t e^{2\lambda_k(t-s)}ds\right)I_2 = \frac{\sigma_k^2(e^{2 \lambda_k t} - 1)}{2\lambda_k}I_2 \equiv v_k(t) I_2
   $$
where $I_2$ denotes the $2\times 2$ identity matrix.   Write $a_k(0) = \alpha_k+ i \beta_k$.  Then the real part of $a_k$ is normal with mean $e^{\lambda_k t}\alpha_k$ and variance $v_k(t)$, and the imaginary part of $a_k(t)$ is normal with mean $e^{\lambda_k t}\beta_k$ and variance $v_k(t)$, and the real and imaginary parts are independent.  Then \cite{Gard0x}
 \begin{equation}\label{EAkt}
 \begin{split}
 &\E A_k(t)^2 = \E|a_k(t)|^2 = e^{2\lambda_k t}(\alpha_k^2+\beta_k^2) + 2 v_k^2 =\\  
&e^{2\lambda_k t}A_k(0)^2 + \frac{\sigma_k^2(e^{2 \lambda_k t} - 1)}{\lambda_k}.
\end{split}
\end{equation}
This formula is valid for finite $t$ regardless of the sign of $\lambda_k$.  Suppose that $\lambda_k > 0$ (that is, $c W(2 \pi k /L) > 1$) for some $k$.  Then $a_k(t)$ will exhibit exponential growth.  We treat seperately the cases where all $\lambda_k <0$ and those where there exist $k$ such that $\lambda_k>0$. 

\subsection{Relationship between $\sigma_k^2$ and $\lambda_k$}  
Suppose that $c W(2 \pi k/L) < 1$ for all $k$, that is all $\lambda_k < 0$.  Then the effect of the initial condition dies away, and each $N_k(t)$ converges to a stationary Ornstein-Uhlenbeck process.  The real and complex (independent) components of $N_k(t)$ each satisfy a scalar equation
     $$
     dZ_t = \lambda_k Z_t dt+ \sigma_k dW_t
      $$
which has stationary distribution $N(0, \sigma_k^2/(-2\lambda_k))$, so that, from \eqref{EAkt},

  \begin{equation}\label{kratio}
  \begin{split}
  & \E A_k(t)^2 \sim \E|N_k(t)|^2 \\
  &= \frac{\sigma_k^2}{-\lambda_k} \\
  &= \frac{\sigma^2 R(2 \pi k/L)}{2L(1-cW(2 \pi k/L))} \\
  &= \frac{\sigma^2}{2L} \frac{R(2 \pi k/L)}{(1-cW(2 \pi k/L))}.
  \end{split}
  \end{equation}
  
When do we see the $k$th mode as dominant in a stationary solution?  We should look for a parameter region where $c W(2\pi k/L) < 1$ and thus $\lambda_k=-1+cW(2\pi k/L)<0$ for all $k$ and also the max value of  $\dfrac{R(2 \pi k/L)}{(1-cW(2 \pi k/L))}$ is noticeably larger than its other values.

Notice that when $\lambda_k=0$ there exists a critical point at which center manifold theory applies \cite{Boxler89, Hutt08}. In this paper we do not address center manifold theory. We first confine ourselves to the study of the case $\lambda_k<0$ for all $k$, where we have an explicit expression for the relationship, \eqref{kratio}, between the noise smoothing and the Mexican Hat influences on the solutions to \eqref{nfes}. We then find that, in spite of the increasing exponential in \eqref{EAkt} when some $\lambda_k>0$, we can still use the ratio in \eqref{kratio} to predict the optimal pairings of the width of the noise smoothing, $\eta$, and the coupling strength, $c$, to obtain a clear but transient pattern in some range of $t$.

\subsection{\label{specnoise}Gaussian Noise Smoother}
As stated earlier, our noise $G(t,x)$ is Gaussian with mean 0 and covariance
\begin{equation*}
\E\left[G(t,x)G(s,y)\right]= \min(s,t) r(x-y).
\end{equation*}

If $Z(t,x)$ is a Gaussian family, Brownian in time and uncorrelated in space, then $\E[Z(t,x)Z(s,y)] = \min(s,t)\delta(x-y)$, and, if $g$ is symmetric, then the smoothed noise
    $$
    G(t,x) = \int_\R Z(t,y)g(x-y)\,dy
    $$
has
   $$\E[G(t,x)G(s,y)] = \min(s,t) r(x-y)
   $$
where $r(x-y) = (g \ast g)(x-y)$.  Here we used the symmetry of $g$.  In particular if $g$ is the density of $N(0,\eta^2)$ then $r$ is the density of $N(0,2\eta^2)$, so that
     $$
     r(x) = \frac{1}{2\eta \sqrt{\pi}} \exp(-x^2/4\eta^2).
     $$
The Fourier transform of $r(x)$ is thus
\begin{equation}\label{Rk}
R(k) = \frac{\sqrt{2}}{\sqrt{\pi}} \exp(-\eta^2 k^2).
\end{equation}

In Section \ref{results} we explore by simulation the effect on the emerging spatial pattern of changing the standard deviation, $\eta$, of the Gaussian noise smoother. We note that in addition to the effect of the noise smoother on the dominant Fourier mode, \eqref{kratio}, there is also a variance reduction effect that is dependent on $\eta$ as well. That is, the variance, $\sigma^2_s$, of the smoothed noise for each process, $a_k$, is approximated by 
\begin{equation}\label{varreduc}
\sigma_s^2\approx \sigma ^2\int_{-\infty}^{+\infty}\bigg[\frac{\exp(-x^2/2\eta^2)}{\eta\sqrt{2\pi}}\bigg]^2 dx\\
=\frac{\sigma ^2}{2\eta\sqrt{\pi}}=\frac{\sigma ^2}{3.54\eta}.
\end{equation}
This implies that when $\eta>1/(2\sqrt{\pi})=0.28$, $\sigma_s^2 < \sigma^2$. We should thus expect that smoothed noise might be more effective at revealing spatial patterns than would i.i.d. noise alone, at least when $\eta>0.28$ (c.f., \cite{Nixon08}).

\subsection{\label{wavenum}Mexican Hat Coupling Operator}
We chose a form for $w(x)$, the difference of the two Gaussian functions, that expresses the common biological observation that there is excitation within a small neighbourhood around each location, and inhibition in a somewhat larger neighbourhood around the excitation. Another way to achieve this effect is to multiply by a Gaussian function and then operate with a Laplacian \cite{Marr82}; yet another alternative operator that involves a squared Laplacian is used in \cite{Scars11}. The existence and identity of excitable spatial modes that lead to spatial pattern in neural fields has been studied also using other approaches (see, e.g., \cite{Hutt2003} and for a review see \cite{Bressloff2012}).

To be explicit, the Mexican Hat operator \cite{Murray89} is defined as 
\begin{equation}\label{MH}
w(x) = b_1 \exp(-(x/d_1)^2) - b_2 \exp(-(x/d_2)^2)
\end{equation}
and its Fourier transform is
\begin{equation}\label{Wk}
W(k)=\sqrt{\pi}\Bigg[b_1d_1\exp\bigg[-\frac{(d_1k)^2}{4}\bigg]-b_2d_2\exp\bigg[-\frac{(d_2k)^2}{4}\bigg]\Bigg].
\end{equation}
So that $w(x)$ has indeed a Mexican Hat shape we need $b_1 /b_2 > 1$ and $d_2 /d_1 > 1$. Under the additional condition that $b_2 d_2 ^3> b_1 d_1^3$, then $k_{max}$, the wave number for which $W(k)$ is largest, is given by
\begin{equation}\label{kmax}
k_{max}=\Bigg(4(d_2^2-d_1^2)^{-1} \ln\Bigg[\frac{b_2}{b_1}\bigg(\frac{d_2}{d_1}\bigg)^3\Bigg]\Bigg)^{0.5}.
\end{equation}

\subsection{Interaction of $k,c,\eta$}
Because of its shape-determining role in (13) we produce graphs, Figure 1, displaying the way in which the expression $\dfrac{R(k)}{(1-cW(k))}$ varies with $k$, $c$ and $\eta$.  We use the Mexican hat operator with $b_2=d_1  =1, b_1=1.1, d_2 = 1.2$.  With these values $k_{max} = 2.026$ and $W(k_{max}) = 0.2134$ so that $-1+cW(k_{max}) < 0$ for all $c < 4.685$.   Note that we are treating $\eta$ as a parameter of (14) so that it can indeed take the value of 0.

It can be seen in Figure 1 that when $\eta = 0$ the value $\dfrac{R(k)}{(1-cW(k))}$ is maximized at $k_{max} = 2.026$ for all values of $c$.  It follows that the dominant mode $k$, that is the integer value of $k$ which maximizes $\dfrac{R(2 \pi k/L)}{(1-c W(2 \pi k/L))}$, is approximately $k_{max}L/(2 \pi) = 0.322L\approx8$ (for the value of $L$ used in our simulations).  As the width $\eta$ of the spatial noise smoother increases, the dominant mode tends to occur at lower values of $k$, indicating fewer spatial cycles on the ring. The decline of the dominant mode with increasing noise smoothing width occurs for lower values of $\eta$ when we consider lower values of $c$. For the largest values of $\eta$ the dominant mode is at $k = 0$, as mentioned earlier.

\begin{figure*}[!ht]
\begin{center}
\includegraphics[width=6in]{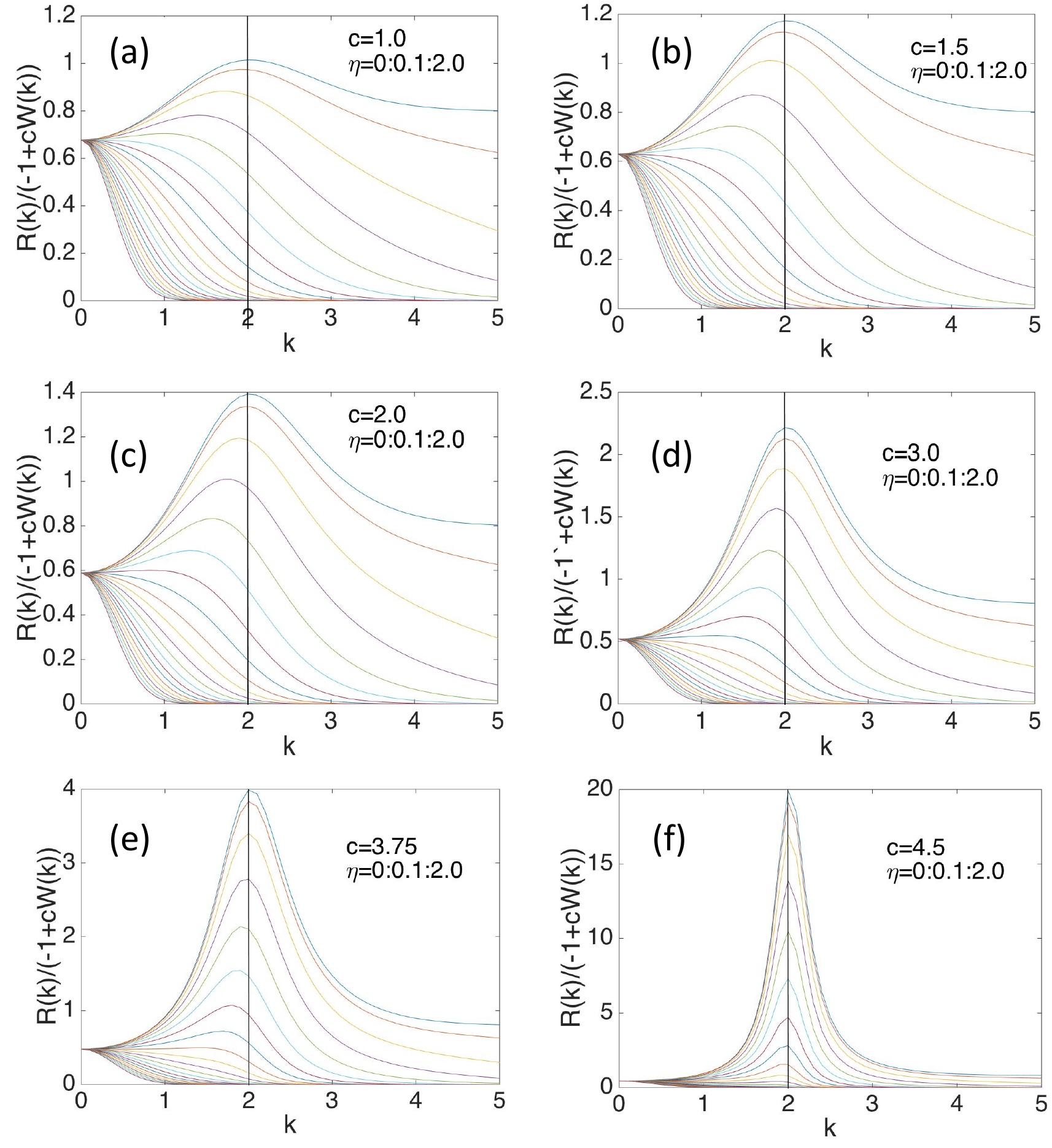} 
\end{center}
\caption{Effects of spatial noise smoother standard deviation, $\eta$, on the ratio $R(k)/(1-cW(k))$ for various values of $c$. Note that the largest values of the ratio occur in the top curve in each plot, where $\eta = 0$, with curves descending in value with increase in $\eta$. Vertical line is at $k = 2$, near the maximum value for all values of $c$ when $\eta = 0$.} 
\label{Figkratio}
\end{figure*}

\section{Simulations}\label{sims} 

An approach consistent with the preceding theory would be to simulate each of the Fourier coefficients $a_k(t)$, given by \eqref{akt}, using the calculations above.  For computational purposes we would simulate only a finite number of them, say $0 \le k \le K$ where $K$ is chosen so that $\E A_k^2$ given by \eqref{kratio} is small for all $k > K$.

In fact we chose to simulate the space-time neural field equation \eqref{nfes} more directly and implemented a discrete (both space and time) approximation using the Euler-Maruyama procedure for obtaining a numerical solution of stochastic difference equations. This approach has been shown to converge rapidly to a close approximation of the continuous solution \cite{Kloeden}.  For a collection of equally spaced points $\{x_j\}$ write $Y(t,x_j) = Y_j(t)$ and $G(t,x_j) = G_j(t)$.  The Euler-Maruyama method gives: 
\begin{equation}\label{discnfes}
\begin{split}
&Y_j (t+\Delta t)-Y_j (t)\\
&= \big[-Y_j(t)+ch \sum_{\ell} w_{j\ell} Y_{\ell}(t)\big]\Delta t+ \sigma  \Bigl(G_j(t+\Delta t)-G_j(t)\Bigr)
\end{split}
\end{equation}
where $w_{j\ell}$ represents the discretized Mexican Hat operator and $h$ denotes the spacing between points ${x_j}$. The factor $h$ is introduced so that the sum in \eqref{discnfes} will be a good approximation to the integral in \eqref{nfes}. 
 
\subsection{\label{simimp}Simulation implementation}
We used $n =128$ points spaced at distance $h = 0.2$ apart in a 1D spatial array, and implemented a periodic boundary so that the values of $x_j$, and the corresponding components $Y_j(t)$ of \eqref{discnfes}, can be thought of as forming a discrete ring of 128 components.  This corresponds to a wrapping length $L = nh = 128 \times  0.2 = 25.6$ in equation \eqref{nfes}.  The 128 components were coupled by the Mexican Hat operator \eqref{MH} with parameter values $b_2 = d_1 = 1$, $b_1 = 1.1$, $d_2 = 1.2$. With these values the function $w(x)$ had effective width from -3 to 3, being very near zero outside this interval, so that our discretized Mexican Hat operator $w_{j\ell}$ was effectively 31 spatial locations wide. That is, the choice $h=0.2$ was convenient for the implementation of the Mexican Hat so that it operated only on 31 of the 128 components around each position in the wrapped 1D spatial array. We chose the width of 31 for the Mexican Hat operator because (a) each single Mexican Hat-type neural coupling in a neural system should span only a fraction of the size of the system, and (b) 31 spatial locations, while being only a fraction of the system size of 128, is large enough to model a Mexican Hat operator that couples several excitatory and inhibitory components. A ``biologically'' based choice would depend on the relative size of an observed neural field to the span of a neighborhood of a neural location that includes both excitatory and inhibitory neighbors.

We implemented space-time noise as described in Section \ref{stnoise}. In the spatially i.i.d.\ case the noise processes $G_j(t)$ were independent standard scalar Brownian motions, while for smoothed noise with parameter $\eta$ the distribution of $G_j(t) = G(t,x_j)$ was as described in Section \ref{specnoise}.  That is, on each iteration independent samples of Gaussian noise for each component were combined for each component by a normalized Gaussian kernel, with standard deviation $\eta$, centered at that component.  In our simulations, presented in Section \ref{results}, we studied spatially i.i.d. noise, and values for $\eta = 0.15, \,0.5,\, 0.67,$ and $1.3$, corresponding roughly to kernels effectively about 1 (i.e., i.i.d., no smoothing), 5, 15, 21, and 39 spatial locations wide. These widths span a range from spatially i.i.d. to a smoothing that combines noise from locations over nearly 1/3 of the total ring.

We solved the stochastic difference equation \eqref{discnfes}, also with $\sigma=0$, iteratively and typically for 10,000 time steps with $\Delta t = 0.00005$, corresponding to a time interval of length 0.5. In a few cases, to be indicated, we extended the simulation to 500,000 iterations and $t=25$, and in a few others we used $\Delta t=0.0025$ in order to reach $t=25$ in only 10,000 iterations. The initial values $Y_j(0)$ for each component were independent random variables chosen uniformly in the interval $[0.5,0.501]$.  The initial perturbation from the constant 0.5 is necessary to observe pattern in the evolving coupled field in the absence of noise \cite{Siebert15}.

For cases where all $\lambda_k<0$ we calculate the amplitude of the dominant harmonic and compare that with the amplitude of this harmonic in simulations. For cases where the maximum $\lambda_k>0$ we illustrate spatial pattern for each set of parameter values for a representative realization of the paths of all 128 components of \eqref{discnfes} with the ring flattened out. We also display for those realizations the Fourier amplitudes (from a Fast Fourier Transform, FFT) of the spatial frequencies as a stochastic process in $t$, and a second measure we term $F$. For both the FFT amplitudes and the computation of $F$ we coarse-grained time, considering 500-iteration time blocks: 1-500, 751-1250, 1751-2250, ...,7751-8250, 8751-9250, 9501-10000.  Note that the gaps between the first two blocks and between the final two blocks are 250 and the gaps between the remaining contiguous timeblocks are all 500. For the FFT amplitudes we averaged Y(t,x) for each component over each 500-iteration block and then computed the FFT on the resulting spatial array. $F$ is a function of the time-block parameter $\tau$ and $\ell$, written as
   \begin{equation}\label{F}
      F(\tau,\ell) = \frac{1}{500} \frac{1}{m} \sum_{s\, \in \,{\rm timeblock} \,\tau} \sum_{j=1}^m |Y_{j+\ell}(s) - Y_j(s)|,
   \end{equation}
where $\ell$ is a spatial offset and $m$ is the distance across the array for which we are computing $F$. In our computations $m$ was fixed at $m = 64$ because the period of the spatial pattern never exceeded this value. In the computation $\ell$ is increased progressively across the spatial array. Thus, whenever the difference $|Y_{j+l}(s)-Y_j(s)|$ is large, the value of $F$ is correspondingly increased. Local maxima in the plot of $F(\tau,\ell)$ occur wherever the spatial offset $\ell$
matches half the period of a spatial pattern. The presence of clear maxima in $F$ indicates the presence of a periodic spatial pattern, and the form of the pattern in $F$ displays the pattern in the $Y_j(t)$ but sometimes more clearly because of the averaging implicit in the computation of $F$.

\section{\label{results}Results}
\subsection{All $\lambda_k<0$}\label{new}
In this section we compare representative calculations based on the theory outlined earlier with simulations of \eqref{discnfes} for cases where all $\lambda_k<0$, and one case where $\lambda_k>0$ for illustrative purposes. In our numerical work, the value of $c$ that separates the case where all $\lambda_k<0$ and where $\lambda_k>0$ for some $k$ is $c\approx 4.685$. 

\subsubsection{No noise}
Before we launch into our study of the joint effects of smoothed noise and the Mexican Hat on the solutions to \eqref{nfes}, we compute the behaviour of \eqref{nfes} without noise.  

The values for the $Y_j(0)$ are i.i.d.\ uniform in $[0.5,0.501]$.  Thus each $Y_j(0)$ has mean value $\mu =0.5005$ and standard deviation $\gamma = 0.001/\sqrt{12} \approx 0.0003$.  Since
    \begin{equation*}
    \begin{split}
    &Y_j(0) = Y(0,jL/n) = \sum_{k=0}^{n-1} a_k(0)e^{2 \pi i k(jL/n)/L}\\
    &=  \sum_{k=0}^{n-1} a_k(0)e^{2 \pi i jk/n}
     \end{split}
     \end{equation*}
we obtain
     $$
     a_k(0) = \frac{1}{n} \sum_{j=0}^{n-1} Y_j(0)e^{-2\pi i jk/n}.
     $$
Taking $k = 0$, we see that $a_0(0)$ is real with mean $\mu$ and variance $\gamma^2/n$.  For $1 \le k \le n-1$ we have $\E[a_k(0)] = 0$ and $\E|a_k(0)|^2 = \gamma^2/n$.   Thus typically $A_k = |a_k(0)|$ is of order $\gamma/\sqrt{n}= (0.001/\sqrt{12})/\sqrt{128} = 0.000026$.

For $c<4.685$, $-1 +cW(2 \pi k/L)<0$ for all $k$ and so no persistent pattern is predicted. Consider, for example, $c=4.5$. The eigenvalue $-1 +4.5W(2 \pi k/L)$ is maximized when $2 \pi k/L \approx k_{max} = 2.0263$, that is for $k = 8$. Thus the eigenvalue for the 8th harmonic is $-1 + 4.5 W( 16 \pi/L) \approx -1+4.5W(k_{max}) = -1 + 4.5 \times 0.22 = -1+0.99 = -0.01$. Over the time interval $[0,0.5]$ the 8th harmonic "grows" (actually damps) by a factor of 
 $$
       e^{(-1+4.5W(16 \pi/L))\times 0.5} \approx e^{-0.005} =  0.995.
       $$
Given the initial amplitude of the 8th harmonic is approximately 0.000026, the typical final amplitude of the 8th harmonic would be 
$$
0.995 \times 0.000026=0.00002587.
$$
We simulated \eqref{discnfes} with $c=4.5$ and the average FFT amplitude of the 8th harmonic over the final 500 iterations (up to t=0.5) from 10 realizations (different realizations of the starting values) was 0.0000227, very close to the predicted value for the dominant 8th harmonic calculated from \eqref{EAkt} with no noise and $c=4.5$. 

To illustrate calculations when $\lambda_k>0$ for some $k$ we now consider a value of $c = 15$.  In this case the eigenvalue for the 8th harmonic is $-1 + 15 W( 16 \pi/L) \approx -1+15W(k_{max}) = -1 + 15 \times 0.22 = -1+3.3 = 2.3$.  Over the time interval $[0,0.5]$ the 8th harmonic in this case does grow by a factor
       $$
       e^{(-1+15W(16 \pi/L))\times 0.5} \approx e^{1.15} =  3.16.
       $$
Since the typical initial amplitude of the 8th harmonic is approximately 0.000026, then the typical final amplitude of the 8th harmonic will be
       $$
       3.16 \times 0.000026  = 0.000082.
       $$

\begin{figure}[!ht]
\begin{center}
\includegraphics[width=3.5in]{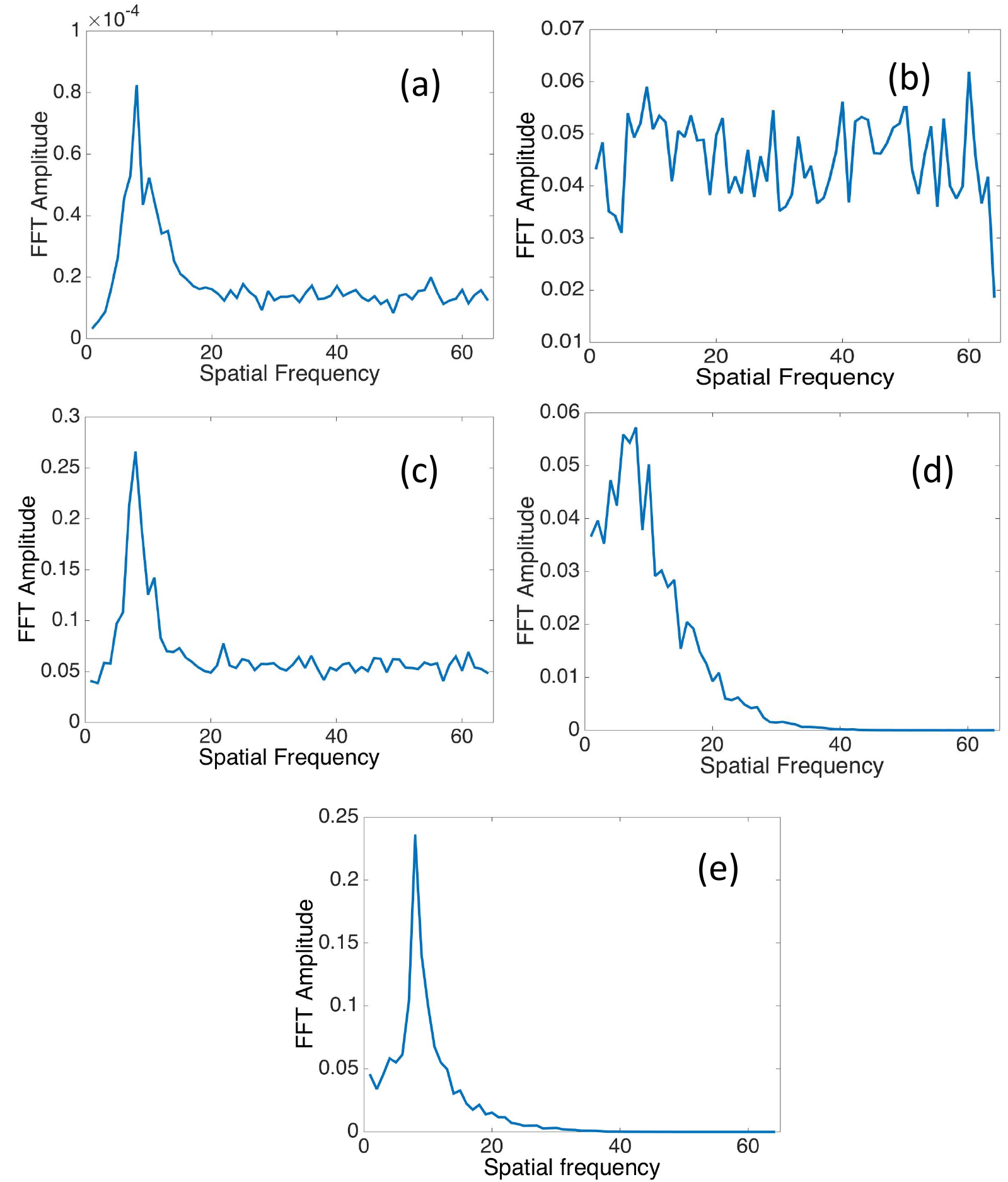} 
\end{center}
\caption{FFT amplitudes averaged over 10 realizations with $\Delta t=0.00005$. (a) No noise, $c=15, \sigma=0,  t=0.5$. (b) I.I.D. noise, $c=4.5, \sigma=1.0, t=0.5$. (c) I.I.D.\ noise, $c=4.5, \sigma=1.0, t=25$. (d) Smoothed noise, $c=4.5, \sigma=1.0, \eta=0.5, t=0.5$. (e) Smoothed noise, $c=4.5, \sigma=1.0, \eta=0.5, t=25$. In all cases the FFT amplitudes are computed on the average $Y_j(t)$ over the final 500 iterations (up to $t=0.5$ or $t=25$), $k_{max}=2.0263$, and the number of cycles in $L/2\pi$ (spatial frequency in graphs) is 8, as described in the text.}
\label{Figklessthan}
\end{figure}
 
We simulated \eqref{discnfes} with $c=15$ and Figure 2a displays the average amplitudes of the spatial harmonics over 10 realizations (different realizations of the starting values). The average value of the FFT amplitude at a spatial frequency of 8 for these 10 realizations of \eqref{discnfes} is 0.000082, exactly the predicted value for the dominant 8th harmonic calculated from \eqref{EAkt} with no noise and $c=15$. 

\subsubsection{I.I.D. noise}
For i.i.d.\ noise the discrete form of the Fourier expansion is valid with $r(x-y)$ replaced by $r_{jk} = \delta_{jk}$.  This in turn gives a version of the Proposition (valid for $x = x_j = jL/n$) with $R(2 \pi k/L)/L$ replaced by $1/n$. As an application of the theory we can use (12) with random initial
values:
   $$
   \E[A_k(t)^2] = e^{2 \lambda_k t} \E[A_k(0)^2] + \frac{\sigma_k^2(e^{2 \lambda_k t}-1)}{\lambda_k}
   $$
where $\sigma_k^2 = \sigma^2/(2n)$ for $k \ge 1$.
     With $c = 4.5$ we have $\lambda_k < 0 $ for all $k \ge 1$, and then
$\E[A_k(t)^2] \to \sigma_k^2/(- \lambda_k)$ as $t \to\infty$.  With $\sigma = 1$ the limiting value $\sigma_k^2/(-\lambda_k)$ is maximized when $k = 8$ and has max value $0.092$.  This corresponds to $A_8(t)$ of order $\sqrt{0.092} = 0.30$, for large $t$.

But this is the limiting value. The value at time $t = 0.5$ is much smaller.  The first term contributing to $\E[A_k(t)^2]$ is at most $\E[A_k(0)^2] = (0.000026)^2$.  The second term is $\dfrac{\sigma_k^2(e^{\lambda_k}-1)}{\lambda_k}$ which is maximized at $k= 8$ with max value $0.0038$. Together $E[A_k(0.5)^2] \le 0.0039$ for all $k \ge 1$. Thus $A_k(0.5)$ is of order at most $\sqrt{0.0039} \approx 0.06$.

We simulated \eqref{discnfes} with $c=4.5$ and Figure 2b displays the average FFT amplitudes for 10 realizations (different realizations of the starting values and the noise). The average value of the FFT amplitude at a spatial frequency of 8 and time period 11 (average over final 500 iterations up to $t=0.5$) for these 10 realizations of \eqref{discnfes} is 0.059, very close to the predicted value for the dominant 8th harmonic from the theory. Notice, however, that the dominant frequency doesn't stand out very well from the noise at other spatial frequencies in Figure 2b. Thus i.i.d.\ noise doesn't substantially enhance the weak pattern expected from $c=4.5$ at this short time interval. 

To see the effect of i.i.d. noise over a longer time interval we simulated 10 realizations of \eqref{discnfes} with $c=4.5, \Delta t=0.00005$ over a time interval from $t=0$ to $t=25$ (500,000 iterations). Figure 2c shows the average FFT amplitude over the final 500 iterations. The average FFT amplitude at a spatial frequency of 8 is approximately 0.27, close to the predicted value of 0.30 for large $t$ from the theory. In order to observe the stochastic paths of $Y_j(t), F$ and FFT amplitude, we also simulated a single realization of \eqref{discnfes} with $c=4.5$ over a time interval from $t=0$ to $t=25$ but with $\Delta t=0.0025$. Figure \ref{Figt25}a shows the results of this simulation. After the longer time interval the pattern more clearly stands out from the noise, consistent with the FFT amplitude shown in Figure 2c. There is still considerable noise, however, evident in the spatial pattern of $Y_j(t)$.

\subsubsection{Coupling and noise smoothing}
The interaction between smoothing width and coupling strength for $\lambda_k<0$ in regard to dominant modes is described by \eqref{kratio}. As depicted in Figure \ref{Figkratio}, the ratio $\dfrac{R(k)}{(1-cW(k))}$ predicts the dominant Fourier mode for various values of $\eta$ and $c$. Noticeably, for any value of $c$, as $\eta$ increases the dominant mode eventually goes to 0. Also, noticeably, for higher values of $c$ this approach to 0 occurs at higher values of $\eta$. In addition, the variance reduction property of the spatial noise smoothing, \eqref{varreduc}, operates for values of $\eta>0.28$, whereas for $\eta<0.28$ variance is actually increased by the `smoothing.' Thus, the noise smoothing has two somewhat conflicting effects: decreasing the variance of the noise-revealed spatial pattern induced by the Mexican Hat operator, but also tending to distort it towards lower spatial frequencies. And this tradeoff also depends on the value of the coupling strength; higher values of $c$ allow the variance reduction effect to override the dominant mode reduction. 

To calculate an example of the application of the theory for smoothed noise we can use \eqref{EAkt} with random initial values as in the case of i.i.d.\ noise. With $c = 4.5. \eta = 0.5$, the limiting value $\sigma_k^2/(-\lambda_k)$ has max value 0.14.  This corresponds to $A_8(t)$ of order $\sqrt{0.14} = 0.37$, for large $t$.

But again this is the limiting value.  The value at time $t= 0.5$ is much smaller.  The first term contributing to $\E[A_k(0.5)^2]$ is at most $\E[A_k(0)^2] =  (0.000082)^2$.  The second term is $\dfrac{\sigma_k^2(e^{ \lambda_k} -1)}{\lambda_k}$.  For $\eta = 0.5$ this is maximized at $k= 5$ with max value $0.0083$.  Together $\E[A_k(0.5)^2] \le 0.0084$ for all $k \ge 1$.  Thus $A_k(0.5)$ is of order at most $\sqrt{0.0084} \approx 0.09$.

We simulated \eqref{discnfes} with $c=4.5, \sigma=1.0, \eta=0.5$ and Figure 2d displays the average spatial FFT amplitude results of 10 realizations (different realizations of the starting values and of the noise). The average value of the FFT amplitude at a spatial frequency of 8 and time period 11 (average over final 500 iterations up to $t=0.5$) for these 10 reactions of \eqref{discnfes} is about 0.06, in line with the predicted value from the theory. Here, however, the dominant mode from the coupling stands out better from the surrounding spatial frequencies, except for the lower frequencies where the dominant mode of the smoother resides. The effect of the smoother is to suppress the noise at frequencies higher than that of the dominant mode, rather than to enhance the dominant mode, whose amplitude is about that with i.i.d.\ noise. Note, however, that both with i.i.d.\ noise and smoothed noise, the amplitude of the dominant mode is substantially greater than that with no noise, even for much larger values of $c$ in the no noise case. Thus, the noise amplifies the spatial pattern, and smoothed noise makes the pattern stand out from the noise.

Because for short simulation times and $\lambda_k<0$ the amplitude processes of the Fourier coefficients and $Y(x,t)$ will be small, we also simulated 10 realizations of the case $c=4.5, \eta=0.5, \sigma=1.0$ over the time interval $t=25$ with the same $\Delta t=0.00005$. For $t=25$ the limiting value of the Fourier coefficient of the 8th harmonic for large $t$ is approximately 0.37. Figure 2e displays the average FFT amplitude from the 10 realizations, which is about 0.24, again in line with the theory. 

\begin{figure}[!ht]
\begin{center}
\includegraphics[width=3.5in]{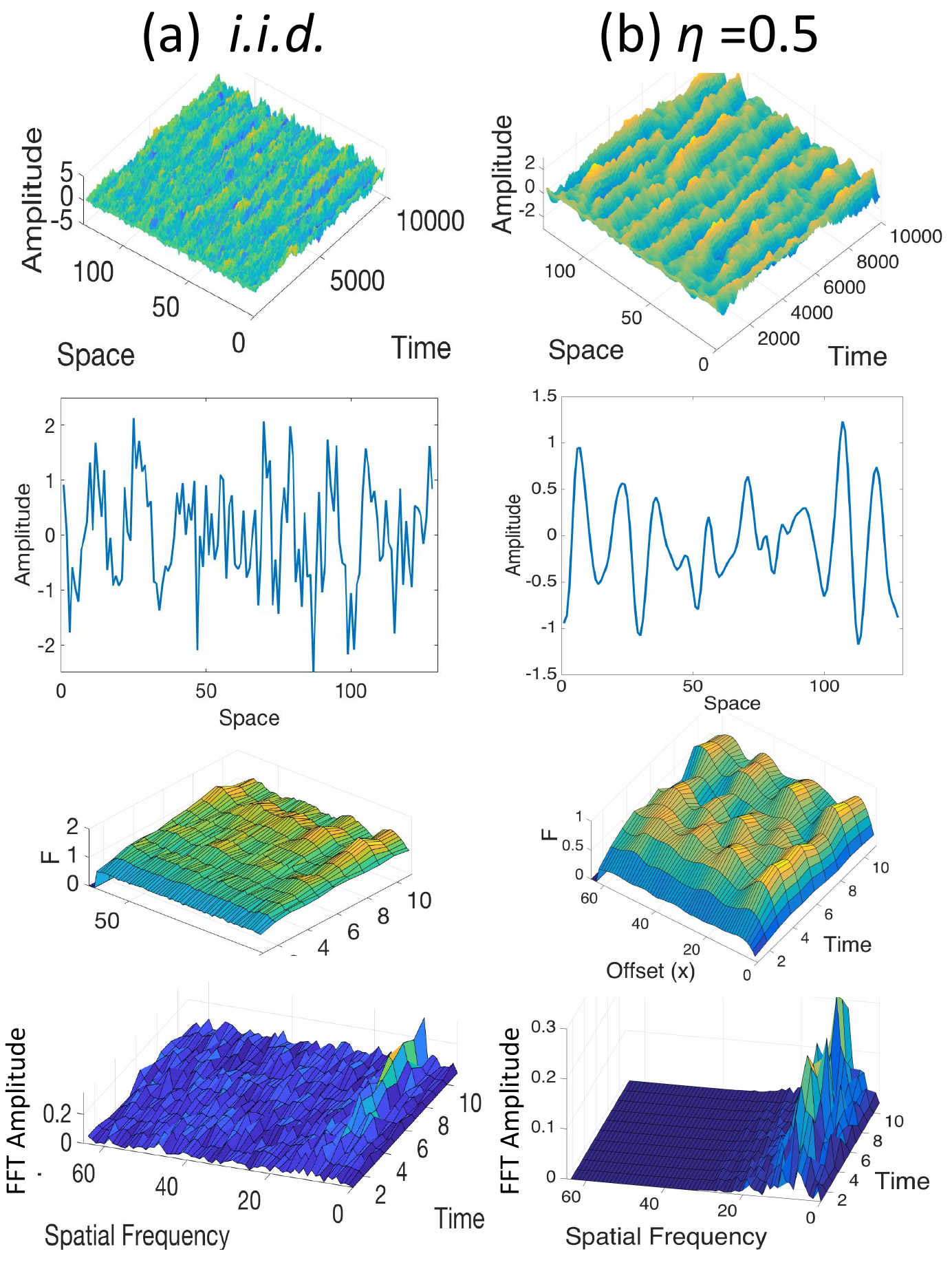} 
\end{center}
\caption{Spatial patterns in the case $\lambda_k<0$ over interval $t=0$ to $t=25$ for i.i.d.\ noise (a) and for smoothed noise with $\eta=0.5$ (b). Here $\Delta t=0.0025, c=4.5, \sigma=1.0$. Top row: amplitude $Y(t,x)$ ; second row $Y(t,x)$ at $t=25$; third row: $F$; bottom row: FFT amplitude. Here the expected number of cycles in $L/2\pi$ is again 8, as described in the text.} 
\label{Figt25}
\end{figure}

In addition, we simulated the case $c=4.5, \eta=0.5, \sigma=1.0$ over the time interval from $t=0$ to $t=25$ but with $\Delta t=0.0025$ so that we could view the sample paths of $Y(t,x), F$ and FFT amplitude. Figure \ref{Figt25}b displays the results of this simulation. The spatial pattern is clearly evident over this longer time interval, again as predicted by the theory. Note that the pattern is quite smooth on the final iteration, in contrast to the pattern with i.i.d.\ noise. Again, the spatial smoothing suppresses the noise at higher frequencies, acting as a bandpass filter, rather than enhancing the spatial pattern itself.

\subsection{Max $\lambda_k>0$}
In this section we display the results of simulations of \eqref{discnfes} with values of $c$ such that $\lambda_k>0$ for some $k$. Here \eqref{EAkt} applies, so we expect exponential increases with time in the amplitudes of the spatial modes, and thus in the spatial pattern. Nonetheless, the interaction of the noise smoothing parameter and the coupling strength are accurately predicted by \eqref{kratio} and Figure \ref{Figkratio}. 
\subsubsection{No noise}
\begin{figure*}[!ht]
\begin{center}
\includegraphics[width=6in]{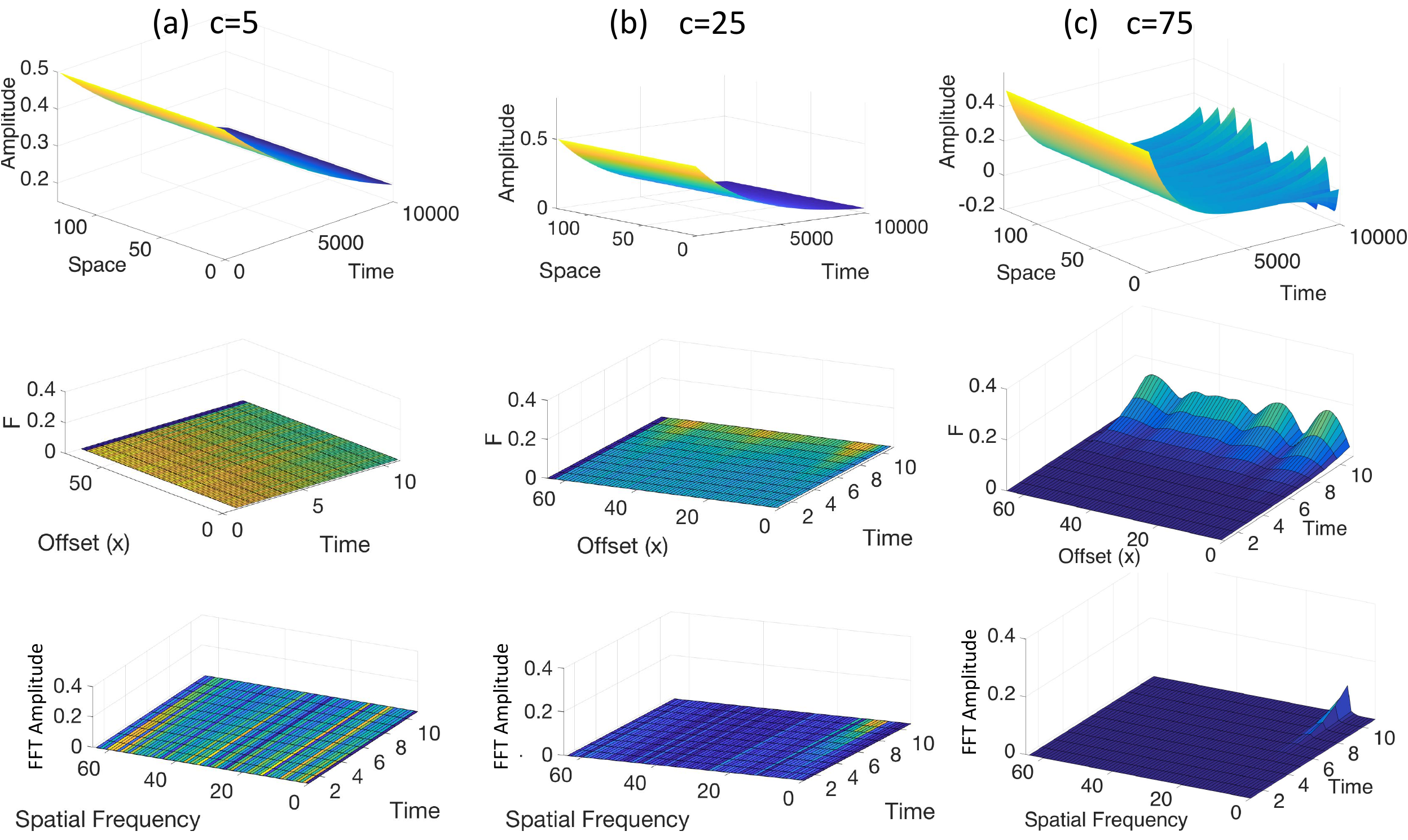} 
\end{center}
\caption{Damping of spatial patterns for which $cW(k_{max})>1$. Top row: amplitude $Y(t,x)$ ; middle row: $F$; bottom row: FFT amplitude. As $c$ (indicated at top of figure) increases the spatial pattern becomes more apparent. Here $\sigma=0$ and thus noise smoothing is irrelevant. Here $k_{max}=2.0263$ and the expected number of cycles in $L/2\pi$ is 8, as described in the text.} 
\label{Figdamping}
\end{figure*}

Figure \ref{Figdamping} displays solutions of \eqref{nfes} with $\sigma=0$, i.e., solutions to the deterministic version of \eqref{nfes} for the 128 spatial replicants. With the parameters $b_2 = d_1 = 1,\,b_1 = 1.1,\, d_2 = 1.2$, as before, we have $W(0) \approx -0.177$ and $k_{max} \approx 2$ and $W(k_{max}) \approx 0.22$.  For $k = 0$ we have an eigenvalue $-1+cW(0) \approx -1 -0.177 c$. This eigenvalue will be negative for all $c$.  When $c = 5$ the most excitable mode has eigenvalue $-1+ 5W(k_{max}) \approx 0.1$, and when $c =25$ the most excitable mode has eigenvalue $-1+25W(k_{max}) =4.5$.  Even though these eigenvalues are greater than 0 we see decay both when $c = 5$ and when $c = 25$, and we do not see the effect of the small initial perturbation away from the constant value $Y = 0.5$ in time $t=0.5$. Here the damping effect is greater than the exponential in \eqref{EAkt} initially, although eventually the exponential causes the pattern to appear (not shown). When $c = 75$, however, the most excitable mode has eigenvalue $-1+ 75W(k_{max}) \approx 15.5$. Moreover, again, $\arg\max\{W(2 \pi k /L: k \in \Z^{\ge 0}\} \approx (L/2 \pi) \times 2 \approx 8$, and so (with no noise) the most excitable mode should have period 8.  By time $t = 0.5$ we begin to see the exponential growth and the pattern of 8 spatial cycles for the $Y_j(t)$ amplitude, FFT amplitude, and $F$, produced by the Mexican Hat operator. See the top graph of Figure \ref{Figdamping}c.

\subsubsection{I.I.D. noise}
\begin{figure*}[!ht]
\begin{center}
\includegraphics[width=6in]{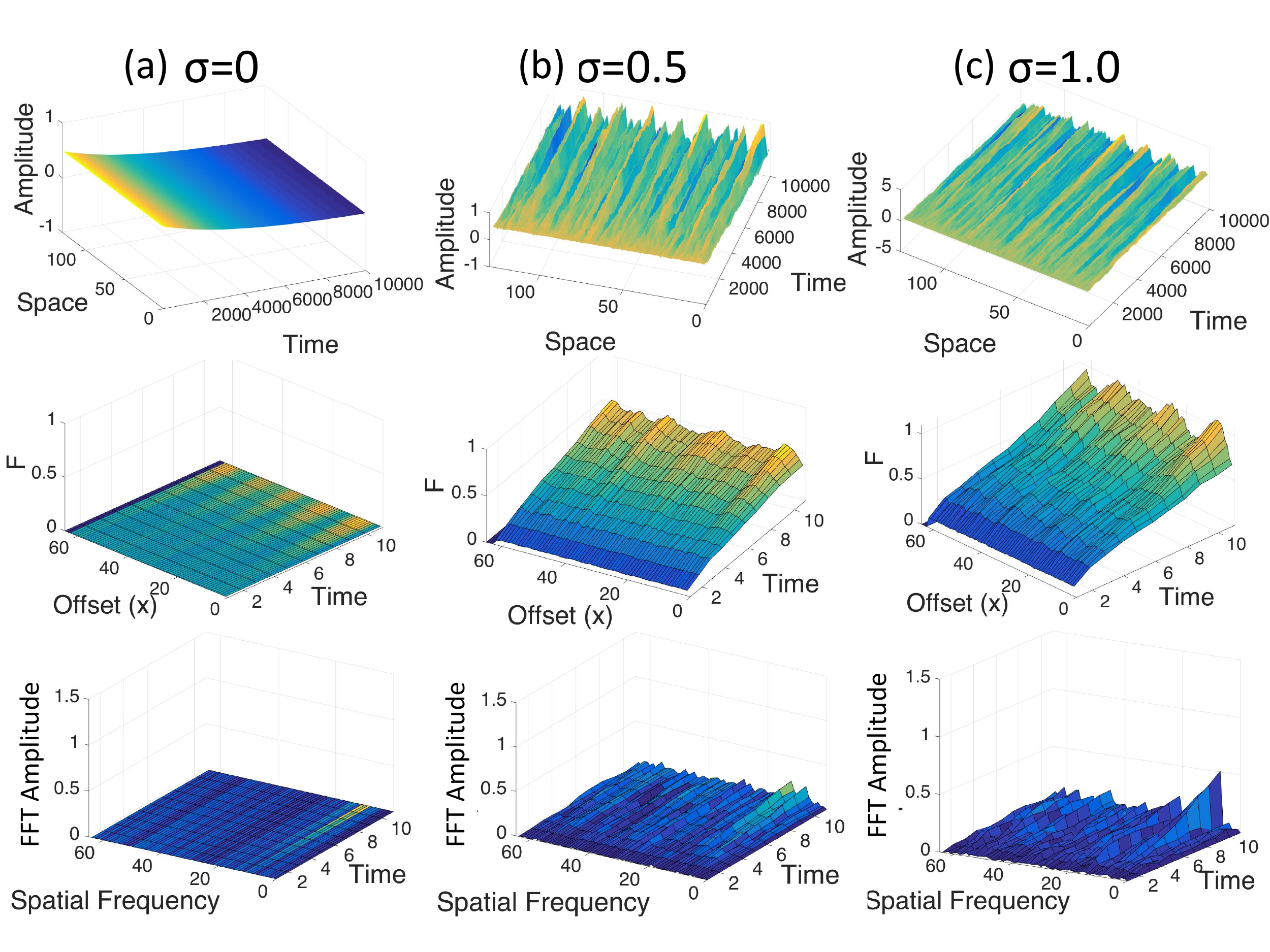} 
\end{center}
\caption{Added noise reveals and sustains the spatial pattern. Rows are the same as in Figure \ref{Figdamping}, noise level $\sigma$ is indicated at the top of each column. Here $c=22.5$, with no noise smoothing.}
\label{FignoiseG}
\end{figure*}

Figure \ref{FignoiseG} displays the effects of adding i.i.d. Gaussian noise to the neural field equation with the same parameters for $w(x)$ as in Figure \ref{Figdamping} except that in Figure \ref{FignoiseG} $c=22.5$ so $\lambda_k>0$ for some $k$. We would expect, based on Figure \ref{Figdamping}b,where $c=25$, that little or no indication of spatial pattern would be apparent when $\sigma=0$, and that is indeed the case (Figure \ref{FignoiseG}a). When a small amount of i.i.d.\ Gaussian noise, $\sigma=0.5$, was added on each iteration in \eqref{discnfes}, however, a spatial pattern is evident (Figure \ref{FignoiseG}b). More noise, $\sigma=1.0$, makes the pattern more apparent (Figure \ref{FignoiseG}c). Thus, again we verify that noise can both reveal weak, otherwise initially (at short time intervals) unobservable, spatial patterns, and also sustain them at observable levels across time. The dominant wave number (in the sense defined earlier) of the spatial pattern does not depend on the standard deviation of the noise, $\sigma$. It will depend, however, on the standard deviation of the smoothing kernel, $\eta$, as predicted by the maximum of $\dfrac{R(k)}{(1-cW(k))}$ in Figure \ref{Figkratio}. We explore this relation in the next section. 
 
\subsubsection{Coupling and noise smoothing}
As discussed earlier, noise smoothing can be expected to affect the spatial pattern created by the Mexican Hat coupling. When the maximum $\lambda_k>0$, as here, \eqref{EAkt} applies. The interaction of $c$ and $\eta$ in \eqref{EAkt} is still controlled by the ratio of $\sigma_k^2/\lambda_k$, so we expect the effects of the noise smoothing to be consistent with those for the case where all $\lambda_k<0$.

\begin{figure*}[!ht]
\begin{center}
\includegraphics[width=6in]{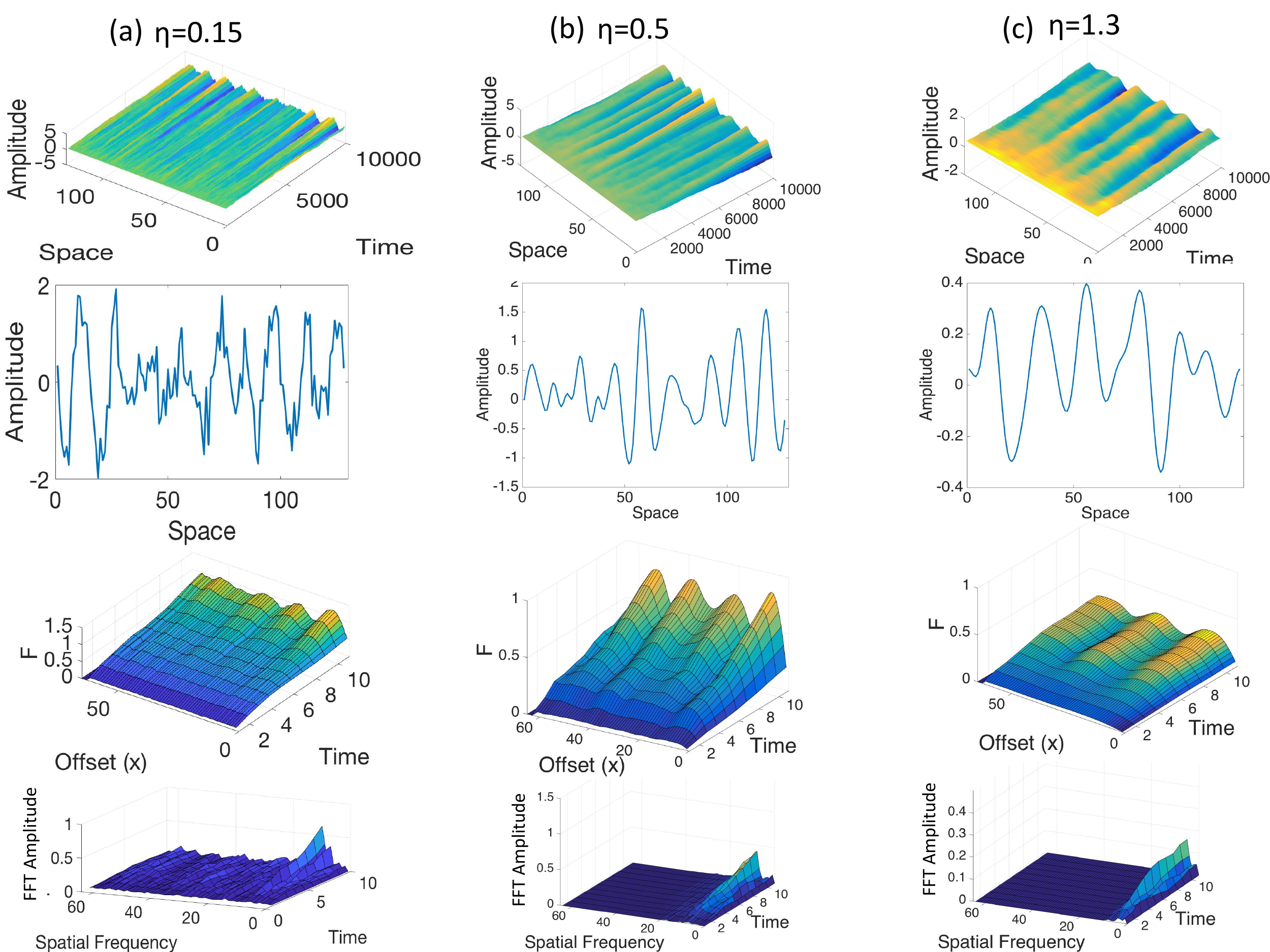} 
\end{center}
\caption{Smoothed noise reveals and sustains the spatial pattern as well as making it more regular. Rows are the same as in Figure \ref{Figdamping} except for addition of row 2, the amplitude after the 10,000th iteration. Smoothing kernel standard deviation, $\eta$, is indicated at the top of each column. Here the standard deviation of the Gaussian noise is $\sigma=1.0$, and the coupling strength is $c=22.5$. The two somewhat contrasting effects of the smoothed noise interact with the spatial pattern created by the Mexican Hat operator.}
\label{Fignsesmooth}
\end{figure*}

These effects are displayed in Figure \ref{Fignsesmooth} for $c=22.5, \sigma=1.0$ and values of $\eta = 0.15, 0.5, 1.3$. Figure \ref{FignoiseG}c shows the results of a simulation with these same parameter values but no noise smoothing. When $\eta=0.15<0.28$, as predicted by \eqref{varreduc}, the variance should be increased slightly but the dominant mode is not affected, and this is apparent in Figure \ref{Fignsesmooth}a. When $\eta=1.3$ the variance is much reduced but the dominant mode is shifted toward 0, reducing by one the number of cycles induced by the Mexican Hat coupling alone (Figure \ref{Fignsesmooth}c). When $\eta=0.5>0.28$, however, the dominant mode is unaffected and the variance also reduced, creating a more apparent spatial pattern, the `best' in our collection of solutions (Figure \ref{Fignsesmooth}b). The interaction between these various factors in their effects on the spatial pattern induced by the Mexican Hat coupling thus creates optimal combinations of $c, \eta$ for the emergence of the spatial pattern. 

This phenomenon of an optimal pair $(c,\eta)$ is reminiscent of stochastic resonance (or stochastic facilitation \cite{McDW11}), in which tuning of the noise strength and threshold yields optimum performance. Of course, the greater $c$ is, the stronger the pattern, so this analogy only applies for situations where damping is sufficient that the pattern is only revealed and sustained by optimally smoothed noise. Smoothing that is too broad imposes a lower frequency on the array, and interferes somewhat with the pattern created by the Mexican Hat operator, as seen in the case where $c=22.5, \eta=1.3$. 

\subsection{\label{smooth}Spatial smoothing of noise without coupling}
\begin{figure*}[!ht]
\begin{center}
\includegraphics[width=6in]{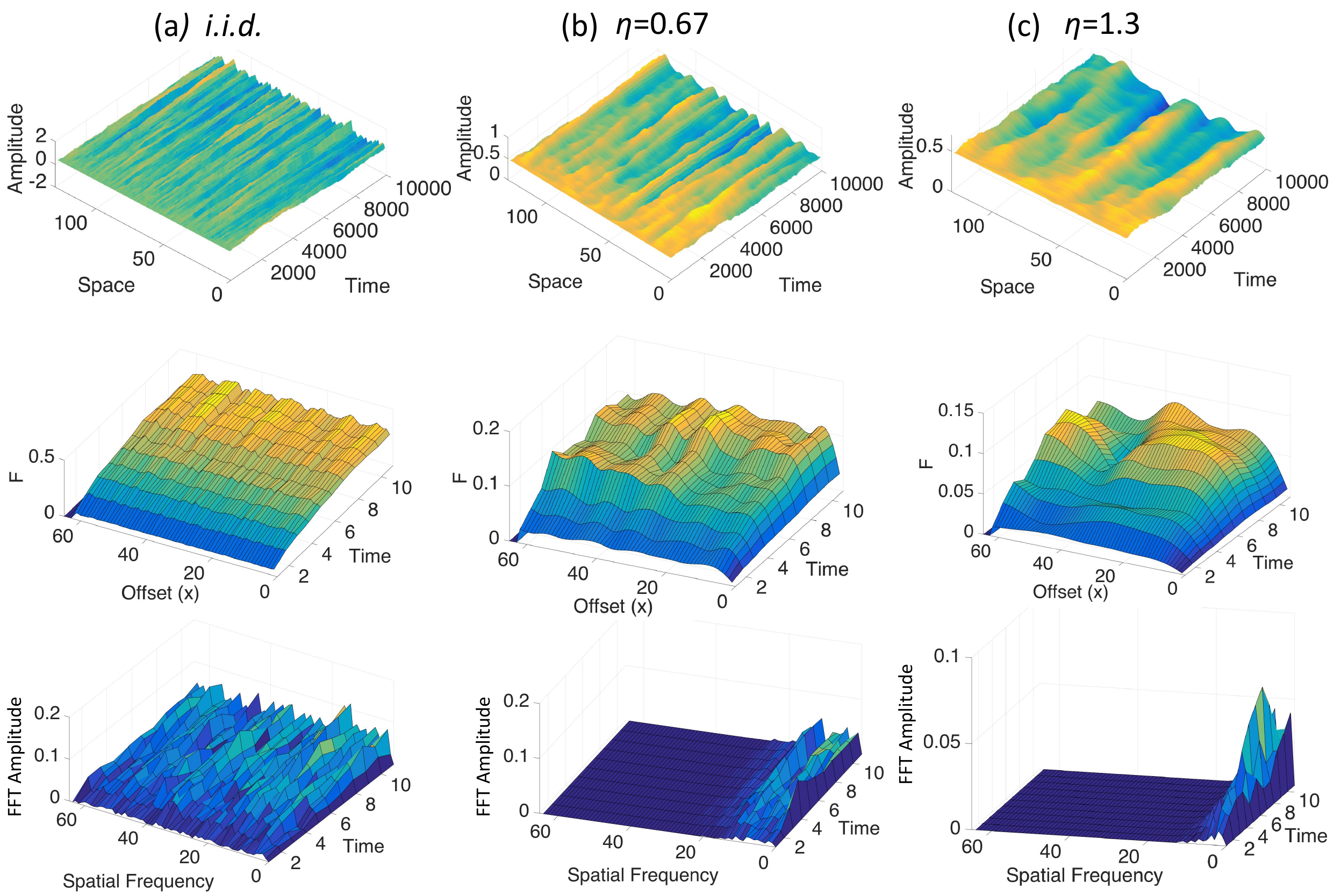} 
\end{center}
\caption{Applying the Gaussian noise-smoothing kernel across neighbors creates a spatial pattern. Rows are the same as in Figure \ref{Figdamping}. Here $c=0$ (so no coupling via Mexican Hat operator), $\sigma=0.5$, and standard deviations of the Gaussian smoothing kernel, $\eta$, are indicated at top of figure and i.i.d. indicates no noise smoothing.}
\label{Figsmoothing}
\end{figure*}

As described earlier, there are spatial modes of the smoothed noise itself. For small $\eta$ these are spread over a range of values of $k$, whereas for large $\eta$ they are concentrated near $k=0$, as shown in Figure \ref{Figkratio}, although the dominant mode is always 0. Because the dominant mode of 0 creates no spatial pattern, however, we might expect that when there are non-zero spatial modes of significant amplitude, the noise alone, in the absence of Mexican Hat coupling, might induce a spatial pattern. Figure \ref{Figsmoothing} displays solutions to the stochastic neural field equation (\ref{nfes}) with $c=0$ and Gaussian-smoothed noise. That is, there is no coupling via the Mexican Hat operator. In these cases, however, $\eta$ has been varied, from i.i.d. spatio-temporal noise to $\eta=0.67$ and to $\eta=1.3$. We observe that smoothing the noise itself creates a spatial pattern with FFT amplitude depending on $\eta$, effecting what we could term a `coupling through partially shared noise.' The FFT amplitude decays exponentially toward the higher frequencies, as expected. With $\eta=2.4$ the FFT amplitude is concentrated close to spatial frequency 0 (not shown). Indeed, when the Gaussian smoother has significant weighting over the entire ring, $\eta>2$, there is only one large bump in the typical simulated path, and the FFT amplitude is concentrated at a spatial frequency of one cycle over $L$ (not shown). Finally, the spatial pattern created by the smoothing kernel can be expected to interact with that created by the Mexican Hat operator to create a sustained and powerful standing wave when the noise smoothing is optimal, as in Figure \ref{Fignsesmooth}b when $\eta=0.5$ or to overwhelm the Mexican Hat pattern when noise smoothing is too great, as in Figure \ref{Fignsesmooth}c when $\eta=1.3$ . 

Pattern produced by Gaussian noise smoothing is, however, significantly different from pattern produced by Mexican Hat coupling. Gaussian noise smoothing with $c=0$ tends to produce stochastic paths that resemble irregular bumps over space, and these bumps move around in the spatial array as the field evolves, making the evolving neural field resemble a chimera of emerging and fading pattern \cite{Kura2002}. The Mexican Hat coupling with or without smoothing tends to result in stochastic paths that resemble stripes rather than bumps (Figures \ref{FignoiseG}, \ref{Fignsesmooth}), although the stripes do move around somewhat as the field evolves, especially when the noise is not optimally smoothed.

\section{\label{disc}Discussion}
We have illustrated, in the context of a "standard" \cite{Faug15} stochastic neural field equation, (\ref{nfes}), that a Mexican Hat convolution kernel produces spatial patterns that (eventually) can be revealed and sustained by noise, even when all eigenvalues are negative, whereas without noise the damping tends to dominate the pattern. Moreover, over long time intervals, Gaussian-smoothed noise alone also produces a spatial pattern, and the two sources of pattern interact. It has been known for some time that a Mexican Hat convolution kernel produces spatial patterns, similar to Turing patterns, in a variety of contexts (e.g., \cite{Murray89,Siebert15}). It was known previously that noise can reveal and sustain such patterns which are otherwise damped, much as quasicycles are revealed and sustained by noise in temporal stochastic processes. We studied major features of this process, exploring the dependence of the pattern on the strength of the Mexican Hat coupling and the width of the noise smoothing. 

First, we found a parametric measure of the interaction between noise sharing/smoothing and Mexican Hat coupling, based on the Fourier expansion of the neural field equation. We then demonstrated, for the case where all eigenvalues of the operator are negative, a close correspndence between the predictions of the continuous theory and an Euler-Maruyama discretization of the theory. Next we explored, for the case where at least one eigenvalue of the operator is positive, the relationship between the coupling strength, $c$ and the width of the noise smoothing, $\eta$. We found a family of `best combinations' of parameters controlling coupling and noise smoothing that produced the strongest patterns. A particular finding is that Gaussian noise smoothing can itself, without coupling, produce spatial patterns in the context of neural field equations, and likely in other contexts as well.

\subsection{Development of spatial pattern in time}
We explored solutions to \eqref{nfes} and \eqref{discnfes} over both short, $t=0.5$, and longer, $t=25$, time intervals. The results over the longer time intervals demonstrate that even when all eigenvalues are negative the Mexican Hat coupling can produce spatial patterns in the presence of added noise, and these patterns can be quite clear when noise is shared/smoothed over a local neighbourhood. The brain, however, does not remain in a single state for such long time intervals. A more likely scenario for application of our results to real brains is that the brain's state changes over shorter time intervals, typically every few hundred milliseconds or faster. Thus, our results for the shorter time intervals are probably most relevant. Importantly, it would be inefficient for the brain to have evolved a system of local coupling, the Mexican Hat coupling, for which all eigenvalues are negative, so that very long time intervals are required for the coupling to create spatial patterns. To illustrate this, compare Figure 2E with Figure 6 middle column. The FFT amplitudes of the spatial patterns are roughly equivalent in the two cases. The time required to reach this state, however, is $t=25$ for Figure 2E, where $c=4.5$ and $\lambda_k<0$ for all $k$, whereas the time required to reach the state in Figure 6 is $t=0.5$, where $c=22.5$ and $\lambda_k>0$ for at least one $k$. Thus, even though we don't know in terms of scaling how this space-time model might relate to real systems, the scenario where at least one $\lambda_k>0$ is the most likely to be present.
Combined with optimal noise sharing, e.g., with $\eta=0.5$ in our simulations and in the example just described, this scenario would implicate a functional role for the local Mexican Hat coupling, such as enhancing edges of neural representations of visual stimuli. 

\subsection{\label{spatcorr}Spatially-smoothed, noise-induced patterns}
The fact that spatially-smoothed noise, i.e. noise that has non-zero correlation length in space, can by itself produce spatial patterns has, we believe, been unappreciated until now. In fact, the Fourier transform of the process \eqref{nfes} predicts its spatial modes, including the case in which, because $c=0$, the Mexican Hat kernel has no effect. In this case, where the smoothing kernel acts alone, a greater spread of noise smoothing leads to a narrower range of $k$ with significant power. Smoothing that is significant over the entire lattice leads to a range of $k$ that is close to 0. The pattern that results from the interaction of the Mexican Hat coupling and the coupling by smoothing of noise is a combination of their respective spatial modes, as reflected in Figure 1. Which modes dominate in a particular implementation depends on the weighting of the respective operators and their extents with respect to the size of the system. 

\subsection{\label{rel}Relation to other work}
There is an extensive literature on stochastic neural field equations. We have chosen a representative, assorted, sample from this literature and point out similarities and differences to the present work, and directions for future studies.

Meyer, Ladenbauer and Obermayer \cite{Meyer17} produced a grid array of spiking neurons with a Mexican Hat coupling structure and measured the covariance pattern of the resulting spike counts while varying the parameters of the Mexican Hat coupling. They found an oscillating pattern of correlation decay around particular fixed neurons, produced by Mexican Hat coupling, with wider inhibitory than excitatory coupling. This pattern did not appear when an inverse Mexican Hat coupling, in which the excitatory coupling extended further than did the inhibitory coupling, was imposed on the grid. The system was driven by external i.i.d. synaptic noise to each neuron, so that the output spike pattern correlation was necessarily produced by the coupling. The approach of the study \cite{Meyer17} was complimentary, or dual, to the approach and objective of the present paper. 

In \cite{Hutt07, Hutt08} there is a cubic reaction term that succeeds in keeping the process stochastically bounded. This is different from the thresholded identity function used in the present work. In \cite{Hutt07} the coupling operator is $(1+(\partial ^2/\partial x^2))^2$, which has an effect similar to a Mexican Hat, whereas in \cite{Hutt08} an effectively Mexican Hat operator, written differently, is used. Both in \cite{Hutt07} and in \cite{Hutt08} the noise is either uncorrelated spatially or, the other extreme, `global fluctuations,' in which the same noise is added to all components of the neural field at each time point. This is in contrast to locally spatially smoothed noise, which we studied here. In these papers the analytic method of center manifold theory together with adiabatic elimination is used to obtain solutions to the neural field equation. 

The paper \cite{Siebert15} studied a model that creates a moving front between states 0 and 1 using a cubic reaction term as in \cite{Hutt07}. At the same time a Mexican Hat kernel together with a diffusion term creates a Turing pattern. Homogeneoous solutions coexist with spatially periodic states. There is no stochastic term, however, and the effects of noise in this model are unknown.

In \cite{Scars11}, Gaussian white noise, as in \cite{Walsh86} together with spatial coupling of the form $(K_o^2+\nabla ^2)^2$ and, again, a cubic reaction term, in a Stratonovich SDE, create patterns in $\mathbb{R}^2$. These patterns resemble various highly regular patterns of vegetation that occur on slopes in semi-deserts around the world.

In Touboul's paper \cite{Touboul14} space-dependent delays are introduced. Again the noise is not smoothed across space. A relevant result is convergence-in-law of network equations. These are in continuous time and discrete in numbers of neurons and of populations, both of which increase to infinity, the `neural-field limit.' The limit is a particular McKean-Vlasov equation, a stochastic neural mean field equation with delays.

These other works suggest various additional ways to pursue the questions studied here. For example, one could insert a cubic reaction term in (\ref{nfes}) instead of using the function $S$ to maintain stochastic boundedness of the process. Additional analytic results might well be obtained using the center manifold theory for the case $k=0$ as in \cite{Hutt07, Hutt08}.

Finally, the present paper can also be seen in the context of the broader field of pattern formation arising from stochastic differential equations. A fairly recent text summarizing many problems and results in this field is that of Bl\"omker \cite{Blom07}. In that work methods are described for the approximation of stochastic partial differential equations near a change of stability using amplitude equations. Bl\"omker focuses on rigorous error estimates for such approximations with the aim of enabling their use in physics and applied mathematics. This text provides many useful clues about how to extend the present work, including a detailed description of approximative center manifold theory.

\begin{appendix}
\section{}
\n {\bf Proposition}: Consider the processes $B_k = \{B_k(t): t \ge 0\}$ defined in \eqref{Bk} for non-negative integers $k$.

(i) The processes $B_0,B_1, B_2, \ldots$ are independent.

(ii) $B_0$ is scalar Brownian motion with variance $L R(0)$.

(iii) For $k \ge 1$, $B_k$ is 2-dimensional Brownian motion with variance $(L/2) R(2 \pi k/L)$.

\n{\bf Proof:} The family $\{B_k(t): k \ge  0, t \ge 0\}$ is a complex valued mean 0 Gaussian process, so it enough to calculate covariances.  Each process $B_k$ is a complex-valued Gaussian process with independent increments in time, so it is enough to calculate covariances of the complex random variables $\{B_k(1): k \ge 0\}.$
Write $B_k(1) = B_k^1+i B_k^2$.  Since $r(x) = r(-x)$ we have
\begin{equation*}
\begin{split}
&\int_0^L (\cos 2 \pi k x/L) r(x)\,dx = R(2 \pi k/L),\\
&\int_0^L (\sin 2 \pi k x/L) r(x)\,dx =0.
\end{split}
\end{equation*}

We calculate some covariances for $k, \ell \ge 0$. First the real parts:
\begin{widetext}
\begin{eqnarray*}
   \lefteqn{\E\left[ B_k^1 B_{\ell}^1\right] }\\
   & = & \E\left[ \left(\int_0^L (\cos 2 \pi k x/L) G(1,x)\,dx\right)\left(\int_0^L (\cos 2 \pi \ell y/L) G(1,y)\,dy\right)\right]\\
    & = &  \int_0^L \int_0^L (\cos 2 \pi k x/L)(\cos 2 \pi \ell y/L) r(x-y) \,dxdy\\
    & = &  \int_0^L \cos \frac{2 \pi k x}{L} \left( \int_0^L \big(\cos \frac{2 \pi \ell x}{L} \cos \frac{2 \pi \ell (y-x)}{L} - \sin \frac{2 \pi \ell x}{L} \sin \frac{2 \pi \ell (y-x)}{L}\bigr) r(x-y) \,dy\right)\,dx \\
    & = &  R(2 \pi \ell/L) \int_0^L (\cos 2 \pi k x/L)( \cos 2 \pi \ell x/L ) \,dx \\
     & = &  R(2 \pi \ell/L)\left\{ \begin{array}{cl} L &\mbox{if } k = \ell  = 0 \\
       L/2 & \mbox{if }\ell  =  k \\
            0 & \mbox{else}.
            \end{array} \right.
     \end{eqnarray*}
Similarly for the imaginary parts:
    \begin{eqnarray*}
   \lefteqn{\E\left[ B_k^2 B_{\ell}^2\right] }\\
   & = & \E\left[ \left(\int_0^L (\sin 2 \pi k x/L) G(1,x)\,dx\right)\left(\int_0^L (\sin 2 \pi \ell y/L) G(1,y)\,dy\right)\right]\\
    & = &  \int_0^L \int_0^L (\sin 2 \pi k x/L)(\sin 2 \pi \ell y/L) r(x-y) \,dxdy\\
    & = &  \int_0^L \sin \frac{2 \pi k x}{L} \left( \int_0^L \big(\sin \frac{2 \pi \ell x}{L} \cos \frac{2 \pi \ell (y-x)}{L} + \cos \frac{2 \pi \ell x}{L} \sin \frac{2 \pi \ell (y-x)}{L}\bigr) r(x-y) \,dy\right)\,dx \\
    & = &  R(2 \pi \ell/L) \int_0^L (\sin 2 \pi k x/L)( \sin 2 \pi \ell x/L ) \,dx \\
     & = &  R(2 \pi \ell/L)\left\{ \begin{array}{cl}
       L/2 & \mbox{if }\ell  =  k  \neq 0\\
            0 & \mbox{else}.
            \end{array} \right.
     \end{eqnarray*}

Finally for the `mixed' terms:
  \begin{eqnarray*}
   \lefteqn{\E\left[ B_k^1 B_{\ell}^2\right] }\\
   & = & \E\left[ \left(\int_0^L (\cos 2 \pi k x/L) G(1,x)\,dx\right)\left(\int_0^L (\sin 2 \pi \ell y/L) G(1,y)\,dy\right)\right]\\
    & = &  \int_0^L \int_0^L (\cos 2 \pi k x/L)(\sin 2 \pi \ell y/L) r(x-y) \,dxdy\\
    & = &  \int_0^L \cos \frac{2 \pi k x}{L} \left( \int_0^L \big(\sin \frac{2 \pi \ell x}{L} \cos \frac{2 \pi \ell (y-x)}{L} + \cos \frac{2 \pi \ell x}{L} \sin \frac{2 \pi \ell (y-x)}{L}\bigr) r(x-y) \,dy\right)\,dx \\
    & = &  R(2 \pi \ell/L) \int_0^L (\cos 2 \pi k x/L)( \sin 2 \pi \ell x/L ) \,dx \\
     & = & 0.
               \end{eqnarray*}
Since orthogonality implies independence for Gaussian random variables, the results follow directly.  \qed
\end{widetext}
\end{appendix}

\section*{Competing interests}
The authors declare that they have no competing interests.
\section*{Author's contributions}
All authors contributed to the conceptualization and writing of the paper. The numerical simulations were accomplished by LMW.
\section*{Acknowledgements}
Lawrence M. Ward was supported by Discovery Grant A9958 from NSERC of Canada.



%

\end{document}